\documentclass[useAMS,usenatbib]{mn2e}

\usepackage{times,graphicx,amssymb,natbib}

\title[Irradiation, The Jeans Mass and Disc Fragmentation]{The Effect of
  Irradiation on the Jeans Mass in Fragmenting Self-Gravitating
  Protostellar Discs}
\author[Duncan Forgan and Ken Rice]{Duncan Forgan $^{1}$\thanks{E-mail:
dhf@roe.ac.uk} and Ken Rice$^{1}$ \\
$^{1}$Scottish Universities Physics Alliance (SUPA), Institute for Astronomy, University of Edinburgh, Blackford Hill, Edinburgh, EH9 3HJ, Scotland, UK}
\begin{document}

\date{Accepted}

\pagerange{\pageref{firstpage}--\pageref{lastpage}} \pubyear{}

\maketitle

\label{firstpage}

\begin{abstract}

\noindent When a self-gravitating disc is subject to irradiation, its
propensity to fragmentation will be affected. The strength of
self-gravitating disc stresses is expected to dictate disc
fragmentation: as the strength of these torques typically decrease
with increasing sound speed, it is reasonable to assume, to
first-order, that disc fragmentation is suppressed when compared to
the non-irradiated case, although previous work has shown that the details
are complicated by the source of the irradiation. We expand on
previous analysis of the Jeans mass inside spiral structures in
self-gravitating discs, incorporating the effects of stellar
irradiation and background irradiation. If irradiation is present,
fragmentation is suppressed for marginally unstable discs at low
accretion rates (compared to the no-irradiation case), but these lower
accretion rates correspond to higher mass discs.  Fragmentation can
still occur for high accretion rates, but is consequently suppressed
at lower disc surface densities, and the subsequent Jeans mass is
boosted.  These results further bolster the consensus that, without
subsequent fragment disruption or mass loss, the gravitational
instability is more likely to form brown dwarfs and low-mass stars
than gas giant planets.

\end{abstract}


\section{Introduction}

\noindent Discs around both supermassive black holes and young
stars are thought to undergo phases in which they
are self-gravitating.  It is possible that if these discs are
sufficiently unstable while they are self-gravitating, they may
fragment into bound objects, providing a mode of planet formation in
protostellar discs \citep{Kuiper1951,Cameron1978,Boss_science} or
star formation in AGN discs \citep{Levin2003,Nayakshin2005}.

The conditions under which a self-gravitating disc is expected to
fragment have been extensively studied.  The most important
criterion is the Toomre parameter \citep{Toomre_1964}

\begin{equation} Q = \frac{c_s \kappa_e}{\pi G \Sigma} <
  1.5-1.7, \end{equation}

\noindent where $c_s$ is the local sound speed, $\kappa_e$ is the local
epicyclic frequency (equal to the angular velocity $\Omega$ in
Keplerian discs) and $\Sigma$ is the disc surface density.  This is a
linear stability criterion - the critical values given above apply to
non-axisymmetric perturbations (see e.g. \citealt{Durisen_review}).  

Once the disc becomes gravitationally unstable, spiral waves are
excited in the disc, producing stresses and providing heating through
shocks.  This stress can be described as a pseudo-viscosity in some
cases
\citep{Shakura_Sunyaev_73,Balbus1999,Lodato_and_Rice_04,Forgan2011},
allowing a simple description of angular momentum transport in the
disc.  The stress produced by the spiral structure can therefore be
described by a Shakura-Sunyaev turbulent viscosity parameter $\alpha$,
and semi-analytic models of self-gravitating discs can be constructed
at low computational cost
(e.g. \citealt{Rice_and_Armitage_09,Clarke_09}).  This pseudo-viscous
approximation has been shown to fail if the disc is too massive or
geometrically thick \citep{Lodato2005,Forgan2011}, but it remains
useful for moderately massive, geometrically thin discs.

A balance can be struck between the shock heating from the spiral arms
and the local radiative cooling.  Discs which achieve this balance are
described as being marginally stable \citep{Paczynski1978}.  This
allows a relationship between the local cooling time normalised to the
local angular frequency ($\beta_c=t_{\rm cool} \Omega$), and $\alpha$:

\begin{equation} \alpha = \frac{4}{9 \gamma(\gamma-1)\beta_c}. \end{equation} 

For fragmentation to be successful, a density perturbation produced by
a spiral arm should be able to grow until it is gravitationally bound.
Therefore, the fragment should be able to cool efficiently to reduce
pressure support due to thermal energy, and continue collapsing.  As
cooling becomes more efficient in a marginally unstable disc (i.e. as
$\beta_c$ decreases), $\alpha$ must increase to redress the balance.
A consensus has developed in recent years that there is a maximum
value of $\alpha$ that the disc can sustain before quasi-steady
self-gravitating torques saturate: $\alpha_{crit}\sim 0.06$
\citep{Gammie,Rice_et_al_05}.  This is supported by numerical
experiments \citep{Cossins2008} that confirm an inverse relationship
between $\beta_c$ and surface density perturbations:

   \begin{equation} <\frac{  \Delta \Sigma_{\rm rms}}{\Sigma}> \propto
               \frac{1}{\sqrt{\beta_c}} \propto \sqrt{\alpha}. \end{equation}

\noindent Once the disc reaches a state such that $\alpha$ cannot
increase to maintain thermal equilibrium, $\Delta \Sigma/ \Sigma$ is
typically of order unity (i.e. density perturbations become
non-linear) and fragmentation can occur.  Rapid cooling also helps to
prevent destructive fragment-fragment collisions, as the initial
fragment spacing will be such that collisions occur within a few
orbital periods \citep{Shlosman1989}. Rapid cooling can then ensure
that colliding fragments do not become unbound as a result of the
collision. The combination of these criteria assure fragmentation can
only occur in the outer regions of protostellar discs, typically at
radii greater than $r = 30 - 40$ AU \citep{Rafikov_05,
  Matzner_Levin_05, Whit_Stam_06,Mejia_3,Stamatellos2008,
  intro_hybrid, Clarke_09,Vorobyov2010,Forgan2011a}.  In the case of
AGN discs, this critical radius is approximately $r = 0.1 pc$
\citep{Levin2003,Nayakshin2007,Levin2007,Alexander2008}.

This description, however, is incomplete.  As fragmentation is
consigned to the outer regions of self-gravitating discs, the local
temperature is unlikely to be determined purely by gravitational
stresses, and is likely to be governed by irradiation, either from the
central object or from an external bath (e.g. envelope irradiation in
embedded protostellar discs).  In this regime, there is no longer a
uniquely determined relation between $\alpha$ and $\beta_c$, as
irradiation provides an extra term to the disc's energy budget.

It is somewhat intuitive to assume that adding extra heating will push
the disc away from marginal instability by increasing the sound speed,
weakening self-gravity and therefore suppressing
fragmentation. \citet{Matzner_Levin_05} demonstrate this using an
analytic prescription for protostellar disc formation and evolution
from Bonnor-Ebert Spheres, showing that fragmentation is typically inhibited for
orbital periods lower than 20,000 years.  \citet{Rafikov2009} also show
that fragmentation is inhibited at large orbital periods when the disc
accretion rate is low.

\citet{Mejia_4} studied the effect of envelope irradiation in grid
based hydrodynamic simulations with radiative transfer, showing that
envelope irradiation suppresses the higher $m$ spiral modes of the
gravitational instability (a result to some extent predicted by
\citealt{Boss2002}, although they disagree on the role of convection
in resisting this suppresion).

\citet{Stamatellos2008} compare stellar and background
irradiation using smoothed particle hydrodynamics (SPH) simulations.
The background irradiation behaviour is similar to that of
\citet{Mejia_4} - stellar irradiation allows the outer disc regions
($r>30$ AU) to cool sufficiently rapidly to fragment according to the
minimum cooling time criterion, but they no longer satisfy the Toomre $Q$
criterion.

More recently, \citet{Kratter2011} have noted that weakening
self-gravity is not generally an impediment to disc fragmentation if
the full effects of mass infall are considered.  As was shown by
\citet{Stamatellos2008}, the cooling time criterion can be satisfied
quite easily in irradiated discs (although maintaining a low $Q$ is
difficult), and that low mass discs in the irradiation-dominated
regime may be made more susceptible to fragmentation given the correct
infall rate.  This is consistent with \citet{Mejia_4}'s finding that
mild irradiation increases gravitational torques (and therefore
increases the equilibrium mass infall rate).

These results should be compared with local shearing sheet simulations
of irradiated discs \citep{Rice2011}.  Under irradiation, there is no
longer a fixed fragmentation boundary using either $\beta_c$ or
$\alpha$.  Comparing to the non-irradiated case, irradiated discs can
fragment for lower values of $\alpha$, but still requires rapid cooling.

With the straightforward criteria for fragmentation in the
non-irradiated regime becoming less clear when irradiation is
incorporated, it is instructive to consider other, more generalised
fragmentation criteria.  We have developed such a criterion, based on
the local Jeans mass inside a spiral wave perturbation
\citep{Forgan2011a}.  If a fragment is to collapse and become bound,
it must have a mass greater than the local Jeans mass.  By measuring
how the Jeans mass evolves with time, we can ascertain whether regions
of a self-gravitating disc are becoming more or less susceptible to
fragmentation.  From this criterion, we are able to estimate not only
when a disc fragments, but what the initial mass of the fragment
should be, an important initial condition to models such as the
``tidal downsizing'' hypothesis of terrestrial and giant planet
formation \citep{Boley2010b,Michael2011,
  Nayakshin2010b,Nayakshin2010,Nayakshin2010a}.

This is somewhat similar to other work which has focused on the
relationship between spiral arm width and the local Hill radius
\citep{Rogers2011a}.  These criteria reflect different competing
influences: the Jeans criterion compares self-gravity to local
pressure forces, whereas the Hill criterion compares self-gravity to
local shear.  The advantage of both these criteria is that they are easily
generalised to cases difficult to describe or explain by traditional
minimum cooling time/maximum stress criteria.  

In this paper, we apply
the Jeans criterion to 1D self-gravitating disc models where the
effects of irradiation are accounted for.  We compare these to models
where irradiation is not present, to assess whether irradiation
promotes or inhibits fragmentation.  In section \ref{sec:method}, we
outline the Jeans criterion and how the 1D disc models are
constructed.  In section \ref{sec:results} we describe and discuss the
results from the models, and in section \ref{sec:conclusions} we
summarise the work. 

\section{Method}\label{sec:method}

\subsection{Calculating the Jeans Mass in irradiated spiral arms}

\noindent To calculate the Jeans mass inside a spiral density
perturbation, we adopt the same procedure as described in
\citet{Forgan2011a}.  We assume the Jeans mass is spherical:

\begin{equation} M_J = \frac{4}{3} \pi \left(\frac{\pi c^2_s}{G\rho_{pert}}\right)^{3/2} \rho_{pert} = \frac{4}{3} \pi^{5/2} \frac{c^3_s}{G^{3/2} \rho^{1/2}_{pert}}. \end{equation}

\noindent Under the thin disc approximation

\begin{equation}\rho_{pert} = \Sigma_{pert}/2H = \Sigma\left(1+ \frac{\Delta \Sigma}{\Sigma}\right)/2H.  \end{equation}

\noindent If we assume the disc is marginally stable, we can obtain

\begin{equation} M_J = \frac{4\sqrt{2}\pi^3 }{3G}\frac{Q^{1/2} c^2_s H}{\left(1 + \frac{\Delta \Sigma}{\Sigma}\right)}. \label{eq:mjeans_nobeta}\end{equation}

\noindent In \citet{Forgan2011a}, we used the empirical result of
\citet{Cossins2008} to estimate the fractional amplitude $\Delta \Sigma/\Sigma$:

\begin{equation} <\frac{\Delta \Sigma_{\rm rms}}{\Sigma}> = \frac{1}{\sqrt{\beta_c}}. \end{equation}

\noindent As we are now considering irradiated discs, it is more
appropriate to use the empirical relationship determined by
\citet{Rice2011}:

\begin{equation} <\frac{\Delta \Sigma_{\rm rms}}{\Sigma}> = 4.47 \sqrt{\alpha}. \end{equation}

 \noindent For the disc to be in thermal equilibrium, the radiative
 cooling, viscous heating and irradiation heating must balance.  The
 dimensionless cooling time, $\beta_c$ is modified thus:

\begin{equation} \beta_c = \frac{(\tau+\tau^{-1})\Sigma c^2_s\Omega}
                {\sigma_{SB} (T^4-T^4_{\rm
                    irr})\gamma(\gamma-1)}, \end{equation}

\noindent where $T_{\rm irr}$ represents the temperature of the local
irradiation field.  

\subsection{The Fragmentation Criterion}

\noindent To produce a generalised fragmentation criterion, we
consider the timescale on which the local Jeans mass changes.  We define:

\begin{equation} \Gamma_J = \frac{M_J}{\dot{M}_J} \Omega.\end{equation}

\noindent If this quantity is small and negative, then the local Jeans
mass decreases rapidly, and fragmentation becomes favourable. Equally,
if this quantity is small and positive, the local Jeans mass increases
rapidly, and fragmentation becomes unlikely. 

The value of $\Gamma_J$ estimates the number of local orbital periods
that will elapse before fragmentation becomes likely (if at all).  In
a steady-state disc in perfect thermodynamic equilibrium, $\Gamma_J
\rightarrow \infty$, and fragmentation will never occur.  Discs that
are not in equilibrium will assume non-infinite values of $\Gamma_J$,
and fragmentation is either likely or unlikely.

We assume $-5 < \Gamma_J < 0$ for fragmentation, but this is yet to be
established empirically.  This is equivalent to requiring a region of
the disc to fragment in less than five orbital periods from the moment
of measuring $\Gamma_J$.  Substituting for $H$ in equation
(\ref{eq:mjeans_nobeta}) gives:

\begin{equation} M_J = \frac{4\sqrt{2}\pi^3 }{3G}\frac{Q^{1/2} c^3_s}{\Omega\left(1 + 4.47\sqrt{\alpha}\right)}. \label{eq:mjeans_sigma} \end{equation}

\noindent We can calculate $\dot{M}_J$ using the chain rule:

\begin{equation} \dot{M}_J = \frac{\partial M_J}{\partial
    c_s}\dot{c}_s + \frac{\partial M_J}{\partial \Omega}\dot{\Omega} +
  \frac{\partial M_J}{\partial \alpha}\dot{\alpha} + \frac{\partial
    M_J}{\partial Q} \dot{Q}. \label{eq:MJdot} \end{equation}

\noindent We will assume $\dot{Q} = \dot{\alpha} = \dot{\Omega} = 0$,
and hence 

\begin{equation} \dot{M}_J = M_J 3 \frac{\dot{c}_s}{c_s} \end{equation}

\noindent From equation (22) of \citet{Forgan2011a}, we calculate

\begin{equation} \frac{\dot{c}_s}{c_s} = 1/2\left(\frac{9\alpha
    \gamma(\gamma-1)\Omega}{4} - \frac{1}{t_{\rm cool}}\right), \end{equation}

\noindent and hence derive:

\begin{equation} \Gamma_J = \left(3/2\left(-\frac{1}{\beta_c} + \frac{9\alpha \gamma(\gamma-1)}{4}\right)\right)^{-1}. \label{eq:betaj}\end{equation}

\noindent As in \citet{Forgan2011a}, we assume that the maximum value
for $\alpha$ saturates at some value $\alpha_{\rm sat}=0.1$, which is
slightly higher than the canonical value of 0.06
\citep{Gammie,Rice_et_al_05}.  We take this value as a conservative
estimate given current uncertainties regarding convergence of 3D
simulations of fragmentation \citep{Meru2011,Lodato2011,Rice2012}.  If
the local value of $\alpha < \alpha_{\rm sat}$, then the system is in
thermal equilibrium and $\Gamma_J \rightarrow \infty$.  Once the local
$\alpha$ reaches $\alpha_{\rm sat}$, it is fixed at this value, while
$\beta_{c}$ is allowed to decrease, reducing the local Jeans mass at a
rate given by $\Gamma_J$.  Comparing $\Gamma_J$ to the critical range
of values given earlier then determines whether a fragment shall be formed.

We should note that we do not consider infall in this analysis,
although we do assume the discs reach a steady state where the disc
accretion rate $\dot{M}$ is the appropriate value to process material
to maintain a constant disc mass (i.e. the local rate of change of
surface density $\dot{\Sigma}=0$).  From the above, we could suppose
that a disc experiencing infall such that $\dot{\Sigma}>0$ may give
$\Gamma_J<0$, and as such irradiation could encourage fragmentation
provided that $Q$ does not also increase.  We leave investigation of
this possibility to future work.

It is unclear from this analysis whether irradiation should drive or
stop fragmentation - consequently it is instructive to test this using
semi-analytic disc models.

\subsection{Simple Irradiated Discs with Local Angular Momentum Transport}

\noindent As in \citet{Forgan2011a}, we construct simple disc models
in the same manner as \citet{Levin2007} and \citet{Clarke_09} to
evaluate the dependence of the Jeans mass on disc parameters.  We fix
the disc's accretion rate 

\begin{equation} \dot{M} = \frac{3 \pi \alpha c^2_s \Sigma}{\Omega} \label{eq:mdot}\end{equation}

\noindent to be constant for all radii, and impose an
outer disc radius.  Assuming marginal instability ($Q=2$) and
thermal equilibrium then fixes the disc surface density profile.  If
thermal equilibrium demands $\alpha > \alpha_{\rm sat}$, then $\alpha$
is set to $\alpha_{\rm sat}$, and $\Gamma_J$ can then become a finite,
negative quantity.  If, or when, $-5< \Gamma_J <0$, the disc model is
assumed to fragment, and the local Jeans mass is calculated.

\noindent We fix the star mass at $1 M_{\odot}$, and consider four
scenarios:

\begin{enumerate}
\item No irradiation ($T_{irr}(r)=0 K$),
\item Envelope irradiation, $T_{\rm irr}(r)=10K$ at all radii,
\item Envelope irradiation, $T_{\rm irr}(r) = 30K$ at all radii
\item Irradiation from the central star. 
\end{enumerate}

\noindent For the third case, we use the radiation field
expression given by \citet{Hayashi1981}:

\begin{equation} T_{irr}(r) = 280 K \left(\frac{M_*}{1 M_{\odot}}\right)
  \left(\frac{r}{1 AU}\right)^{-1/2}. \end{equation}

\noindent For $\alpha_{\rm sat}=0.1$ and $\gamma=5/3$, we can actually 
determine the maximum value of $\beta_c$ at which fragmentation
occurs (i.e. where $\Gamma_J = -5$), which we find to be

\begin{equation} \beta_{\rm crit} < 2.2. \end{equation}

\noindent Note the conventional critical cooling time for this value
of $\gamma$ is $\beta_{\rm crit} = 3$.  We have however assumed a
value of $\alpha_{\rm sat}$ around $60\%$ larger than the usual value,
so our slightly smaller value of $\beta_{\rm crit}$ compares sensibly
to the typically used cooling time criteria.  Altering the critical
value of $\Gamma_{J}$ instead would also allow $\beta_{\rm crit}$ to
be increased to a more standard value, but this would permit discs to
fragment on timescales longer than might be considered physical.  In
any case, the resulting fragment masses are only very weakly sensitive
to these considerations.

\section{Results}\label{sec:results}

\subsection{Disc Profiles}

\begin{figure*}
\begin{center}$
\begin{array}{cc}
\includegraphics[scale=0.5]{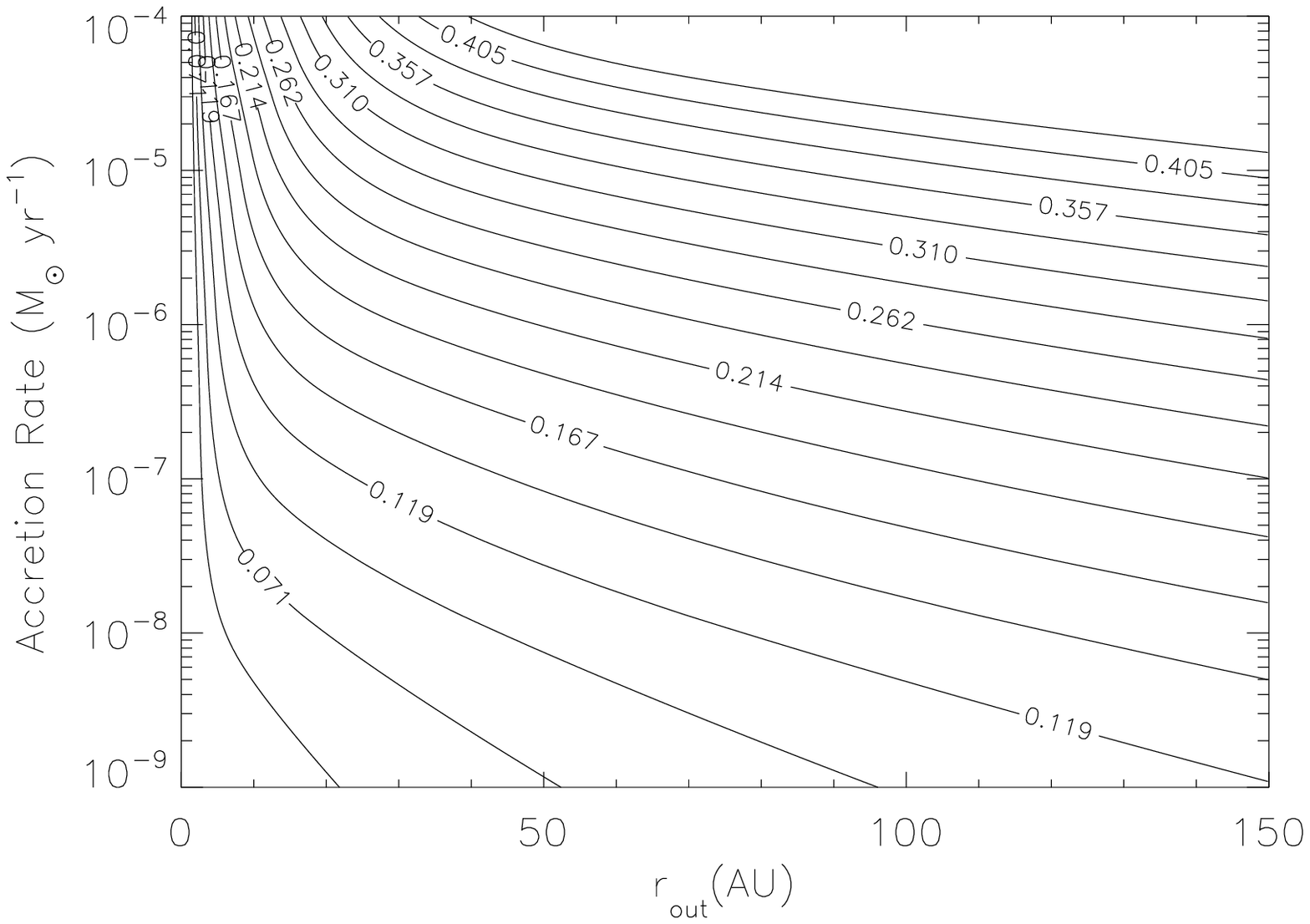} &
\includegraphics[scale = 0.5]{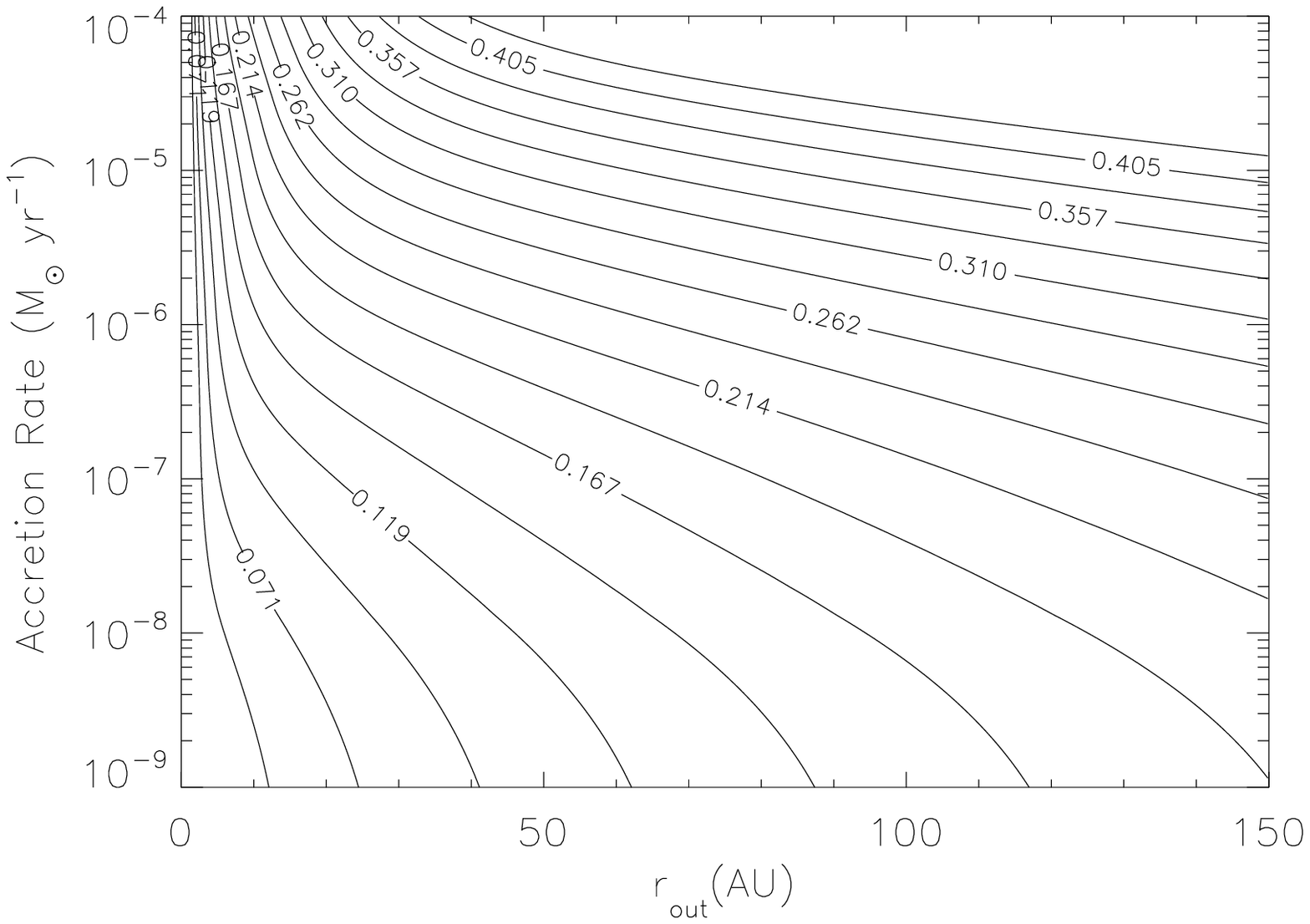} \\
\includegraphics[scale=0.5]{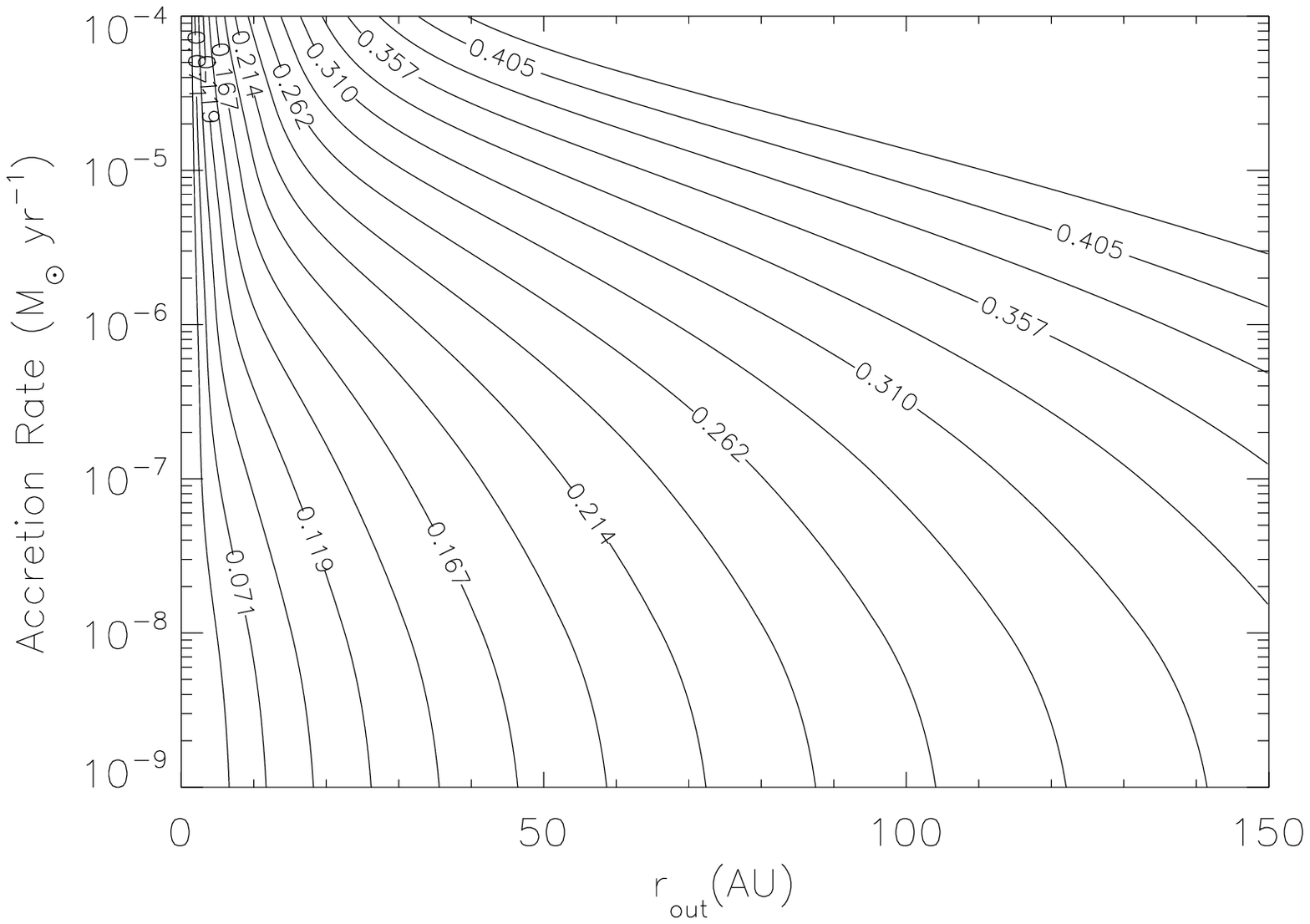} &
\includegraphics[scale = 0.5]{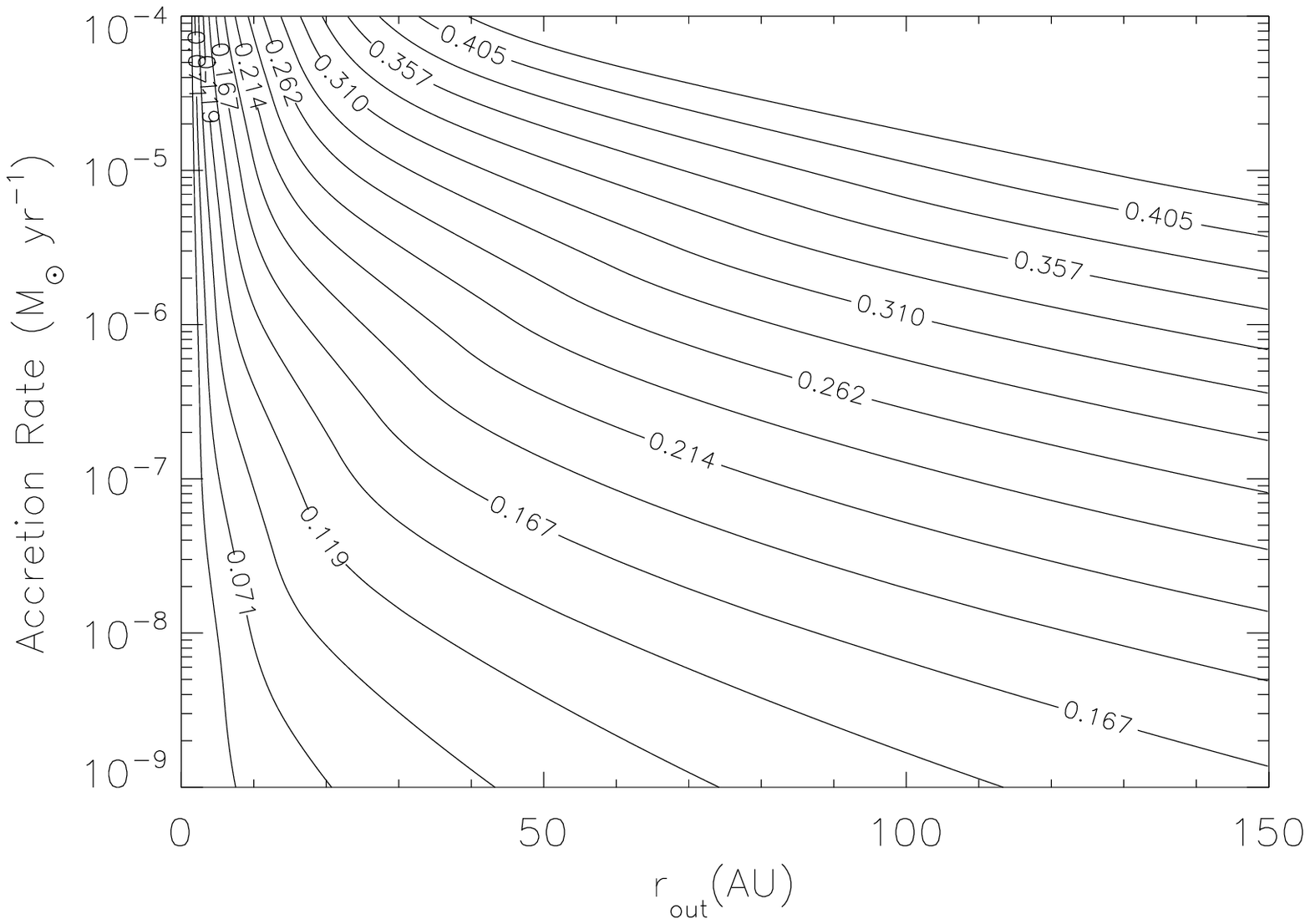} \\
\end{array}$
\caption{2D contour plots of the disc-to-star mass ratio as a
  function of the steady state `pseudo-viscous' accretion rate
  $\dot{M}$, and the disc's outer radius $r_{\rm out}$, for the four cases
  considered in this work: without irradiation (top left), with
  background irradiation at 10K (top right), with background
  irradiation at 30K (bottom left), and with stellar irradiation
  (bottom right).  Note solutions lying in the upper regions
    of the plots are excluded as the disc lifetimes are too short (see
  text).}\label{fig:mdot_r_q}
\end{center}
\end{figure*}

\noindent Figure \ref{fig:mdot_r_q} shows 2D contours of the
disc-to-star mass ratio $q$ in all four scenarios, given the accretion
rate and the disc outer radius.  The differences between the
non-irradiated case (top left in Figure \ref{fig:mdot_r_q}) and the
models discussed in \citet{Forgan2011a} are due to a different
selection of $Q$ (in this paper, we assume $Q=2$, and in the previous
work we assumed $Q=1$), but the same broad features remain: a high
disc-to-star mass ratio is required to maintain a self-gravitating
disc with a modest accretion rate and reasonable outer disc radius.
For example, in the non-irradiated disc, an accretion rate of $10^{-7}
M_{\odot} \mathrm{yr^{-1}}$ and outer radius of $r_{out} = 30$ au
demands a disc mass of 0.119 $M_{\odot}$.

At high accretion rates, the lifetime of material in the disc becomes
comparable to the orbital timescale.  For this reason, we truncate the
upper limit of the contours by demanding that

\begin{equation} \frac{M_{\rm disc}}{\dot{M}} > 5
  \frac{2\pi}{\Omega}, \end{equation}

\noindent which essentially requires the disc to exist for at least
five orbital periods at the given radius.  Any disc model which does not
satisfy this is discarded and not considered in the subsequent
analysis.  This explains the empty regions in the upper right portion
of each plot.

Comparing the non-irradiated case to the other three cases, we can see
that the high $\dot{M}$ section of the parameter space (i.e., above
$\dot{M} \sim 10^{-6} M_{\odot} \mathrm{yr^{-1}}$) remains similar.
The effect of irradiation becomes more apparent at lower accretion
rates, forcing the equilibrium disc structure to be more massive for a
given $\dot{M}-r$ locus. For example, in the case of background
irradiation at 10K, the $0.214 M_{ \odot}$ contour at $r=150$ au moves
downward from $\dot{M}= 10^{-7} M_{\odot} \mathrm{yr^{-1}}$ to $\sim 2
\times 10^{-8} M_{\odot} \mathrm{yr^{-1}}$.  At $T_{\rm irr}=30K$,
this contour intersects the x-axis at $ < 100$ au.  Indeed, discs with
radius 150 au and accretion rates of $10^{-7} M_{\odot}
\mathrm{yr^{-1}}$ have masses around 0.357 $M_{\odot}$.

In the stellar irradiation case, the change is less striking - a
$0.214 M_{\odot}$ disc with radius 150 au is still formed at an
accretion rate of $\sim 2 \times 10^{-8} M_{\odot} \mathrm{yr^{-1}}$,
but the lower $\dot{M}$ region allows higher mass discs to have larger
outer radii than the background irradiated cases.

Adding extra mass to the outer regions of self-gravitating discs will
in general encourage fragmentation
\citep{Vorobyov2010,Kratter2010a,Kratter2011,collapses}, so we might
expect to see the Jeans criterion satisfied at lower accretion rates
in the presence of irradiation.  Conversely, increasing the disc sound
speed will increase the typical fragment masses (if fragmentation can
still be achieved).

\subsection{Jeans masses}

\begin{figure*}
\begin{center}$
\begin{array}{cc}
\includegraphics[scale=0.5]{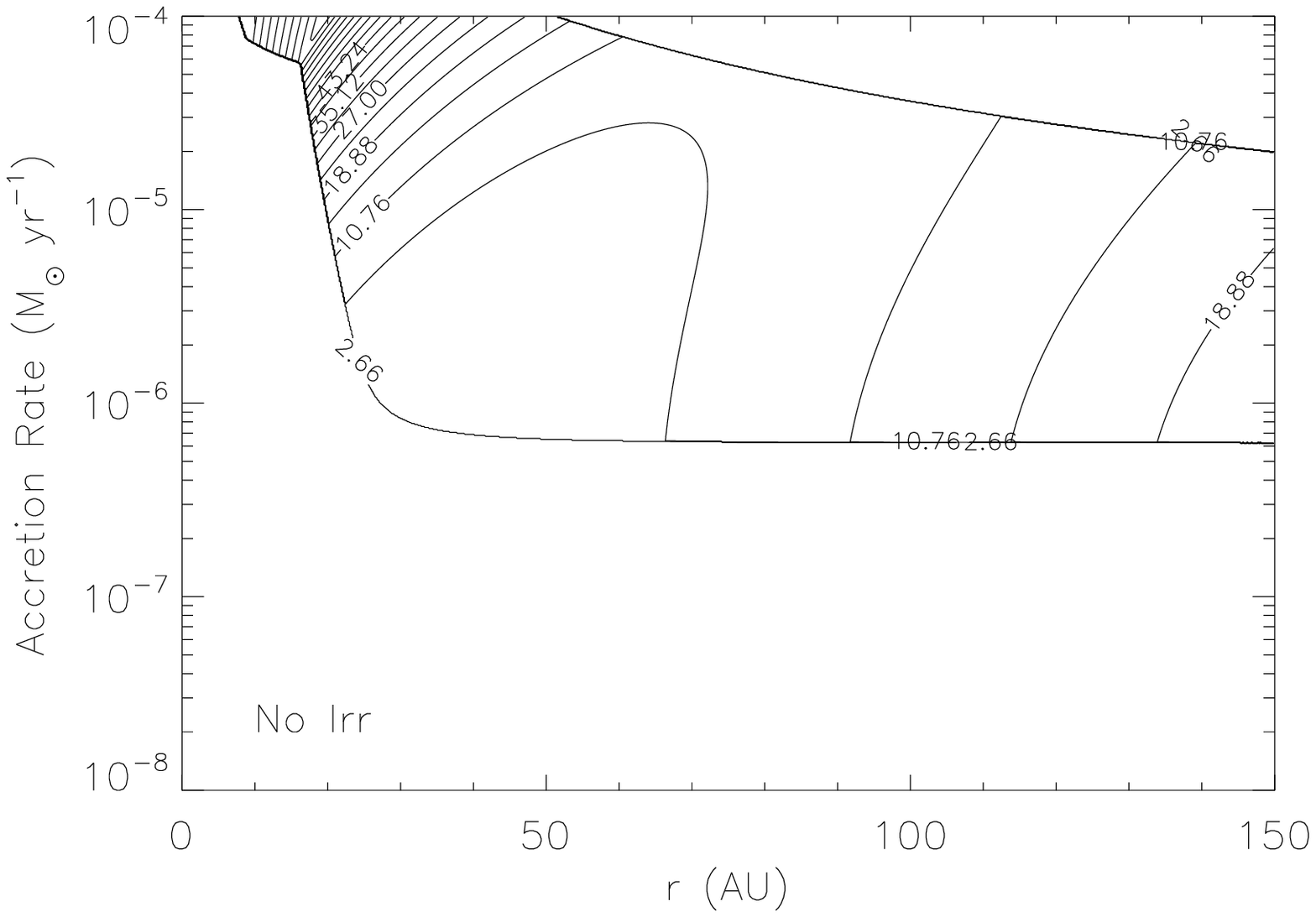} &
\includegraphics[scale=0.5]{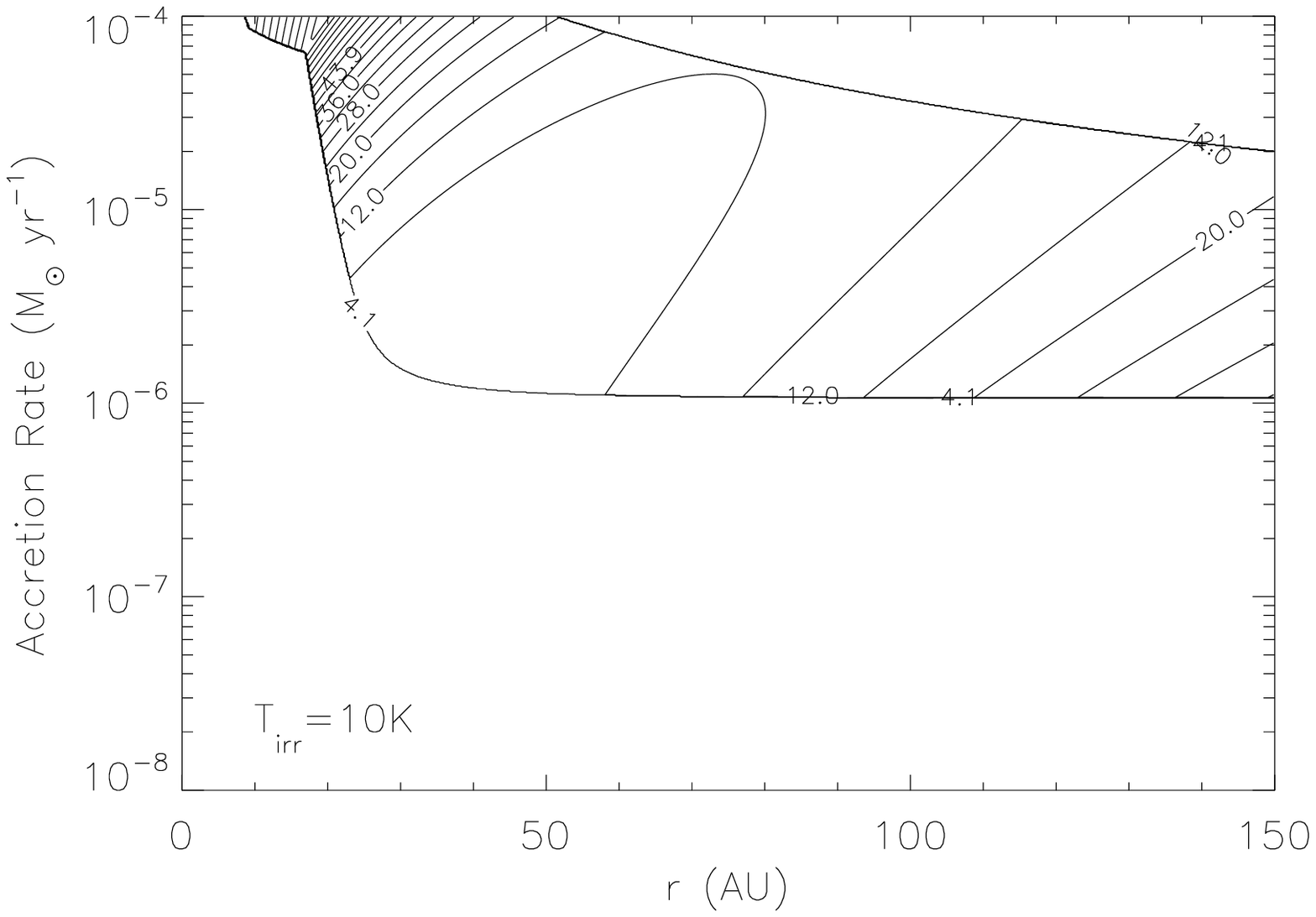} \\
\includegraphics[scale = 0.5]{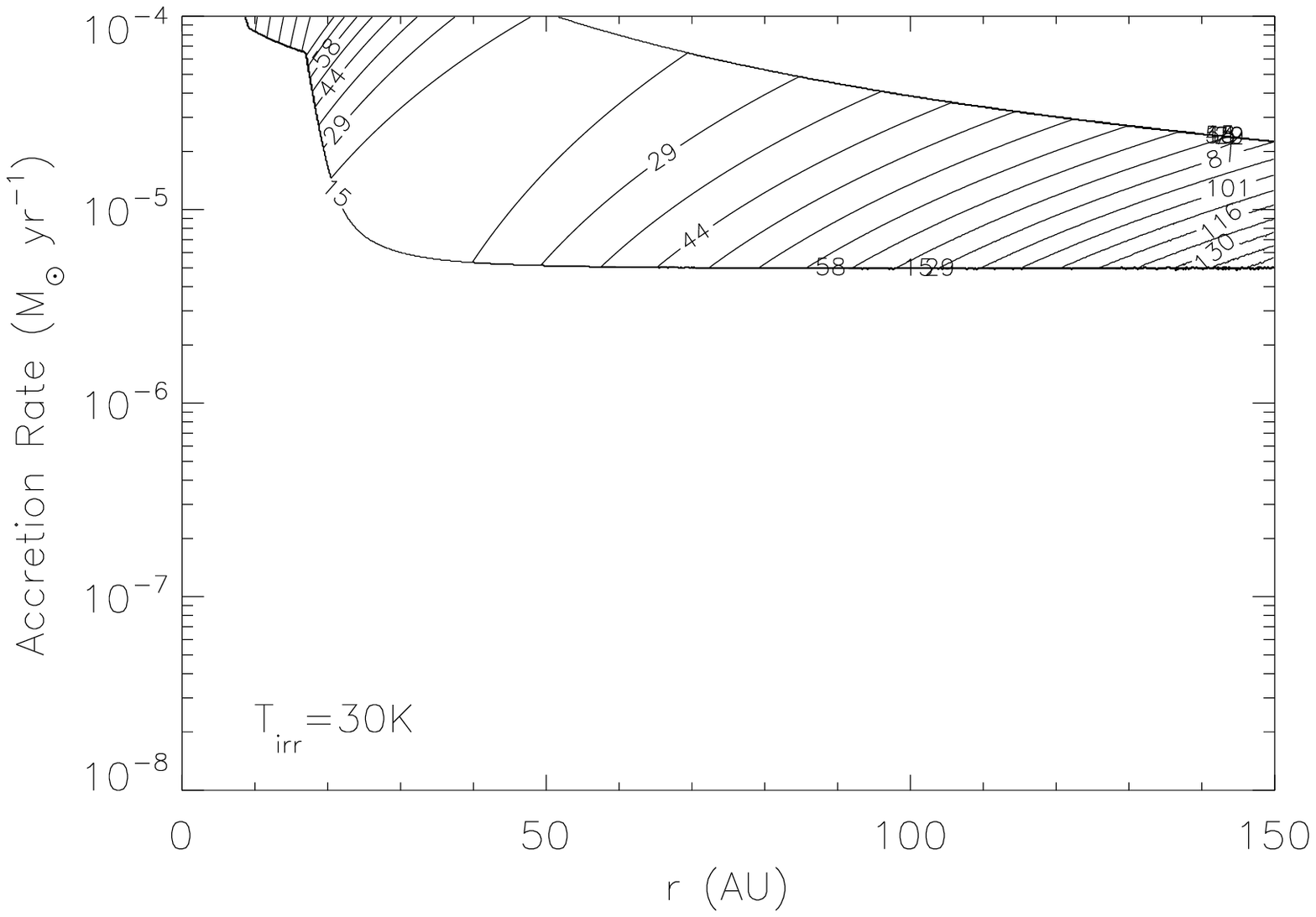} &
\includegraphics[scale = 0.5]{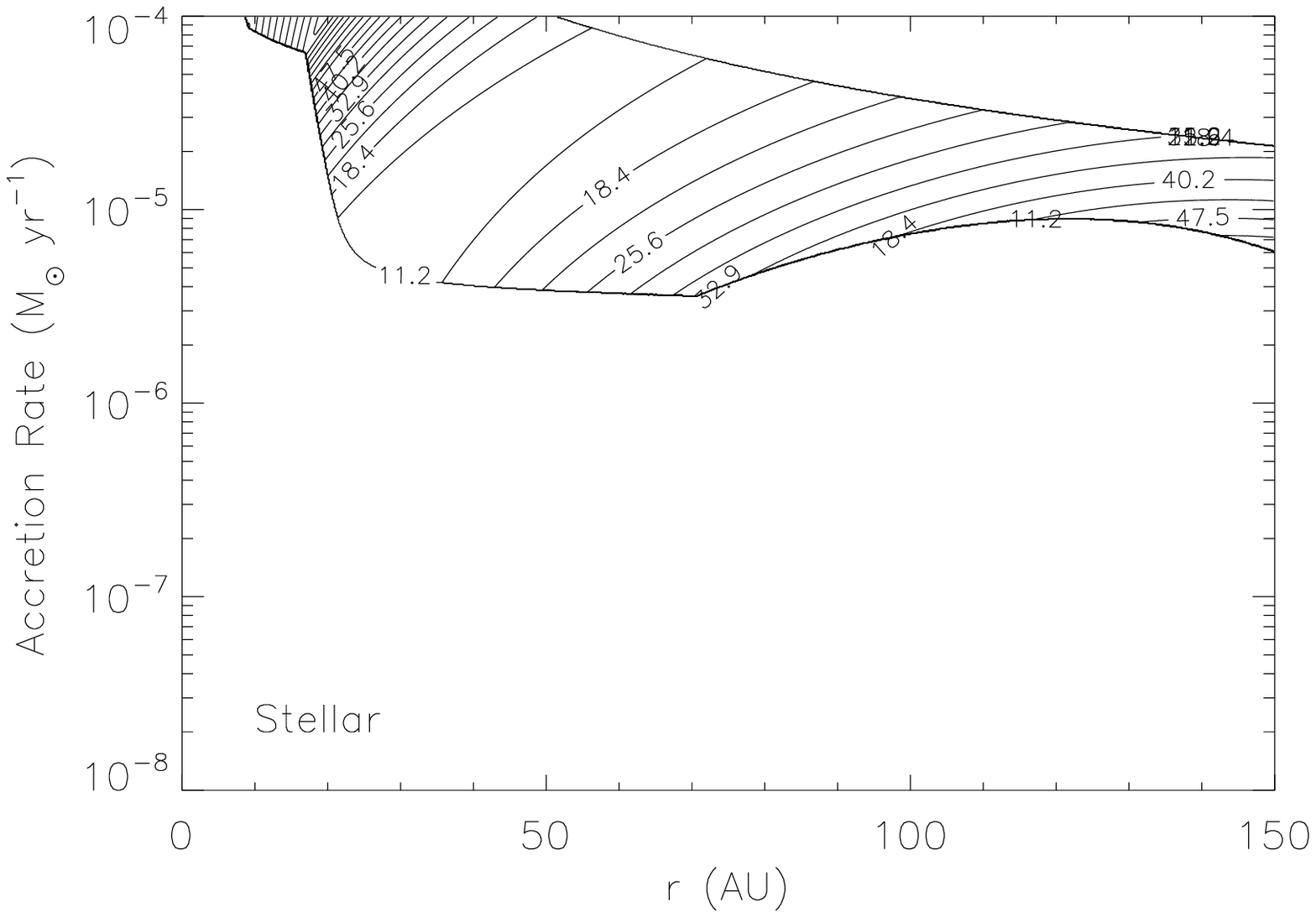} \\
\end{array}$
\caption{2D contour plots of the Jeans mass inside a spiral
  perturbation, as a function of the steady state `pseudo-viscous'
  accretion rate $\dot{M}$ and radius in the disc $r$, for the four
  test cases considered in this work: without irradiation (top left),
  with background irradiation at 10K (top right), with background
  irradiation at 30K (bottom left), and with stellar irradiation
  (bottom right). Again note that the upper regions of the parameter
  space are excluded because the disc lifetimes are too short.
  Regions beneath the lowest contour are marginally unstable, but in
  thermal equilibrium, and do not fragment.}\label{fig:mdot_r_frag}
\end{center}
\end{figure*}

\noindent Figure \ref{fig:mdot_r_frag} shows the regions of parameter
space in which the four scenarios can fragment, as well as the local Jeans
mass at the point of fragmentation.  

The non-irradiated case is qualitatively similar to that found by
\citet{Forgan2011a}, with the minimum Jeans mass also taking a similar
value of 2.66 $M_{\rm Jup}$.  Self-gravitating discs will not fragment
at low disc radii (except at very high, probably unphysical accretion
rates).  There is a minimum accretion rate required for fragmentation
even at large disc radii, corresponding in this case to around $7
\times 10^{-7} \mathrm{M_{\rm \odot} yr^{-1}}$.  By comparison with
Figure \ref{fig:mdot_r_q}, we can see this sets a minimum mass at
which discs can fragment at around $0.2 M_{\rm \odot}$ for a disc
outer radius of 50 au.

At high accretion rates, the regions of parameter space susceptible to
fragmentation remain roughly similar across all four cases, but the
irradiated discs show suppression of fragmentation at lower accretion
rates.  The 10K irradiation case shows fragmentation at accretion
rates around a factor of 1.25 higher than the no irradiation case.
This increases to a factor of 5 as the background temperature is
increased to 30K. The stellar irradiation case displays similar
behaviour at low radii, but at around 70 au we see discs with low
accretion rates possessing sufficiently low optical depths for the
irradiation to heat the midplane so that $T_{\rm irr} \sim T$. While
marginally unstable solutions still exist, fragmentation is suppressed
at low accretion rates.  At radii of $\sim 120$ au, accretion rates as
high as $10^{-5} M_{\rm \odot} \mathrm{yr^{-1}}$ are required to produce a
fragmenting disc.

The minimum Jeans mass is typically larger for the three irradiated
cases: 4.1 $M_{\rm Jup}$ in the $10K$ irradiation case, 15 $M_{\rm Jup}$ in
the $30K$ case and 11.2 $M_{\rm Jup}$ for the stellar irradiation case.
Also, the rate at which the Jeans mass increases with disc radius is
higher than in the non-irradiated case.  This further supports the
creation of low mass binary star systems as opposed to single-star
planetary systems \citep{Clarke_09}.

In short, irradiated discs tend to produce more massive fragments.
There are two reasons for this: 

\begin{itemize}
\item The effective $\alpha$ generated by the instability is reduced
  in the presence of irradiation, and hence at a given $\dot{M}-r$
  locus, the discs are more massive, and
\item The steady state temperatures in irradiated discs are higher,
  providing extra pressure support against gravitational collapse.
\end{itemize}

\noindent Further, if we consider the expected isolation masses of
these fragments \citep{Lissauer1987}, then we expect irradiation to
push fragments well into the brown dwarf/low-mass star regime (see also
\citealt{Kratter2009}).

\subsection{Surface Densities at Fragmentation}

\begin{figure*}
\begin{center}$
\begin{array}{cc}
\includegraphics[scale=0.5]{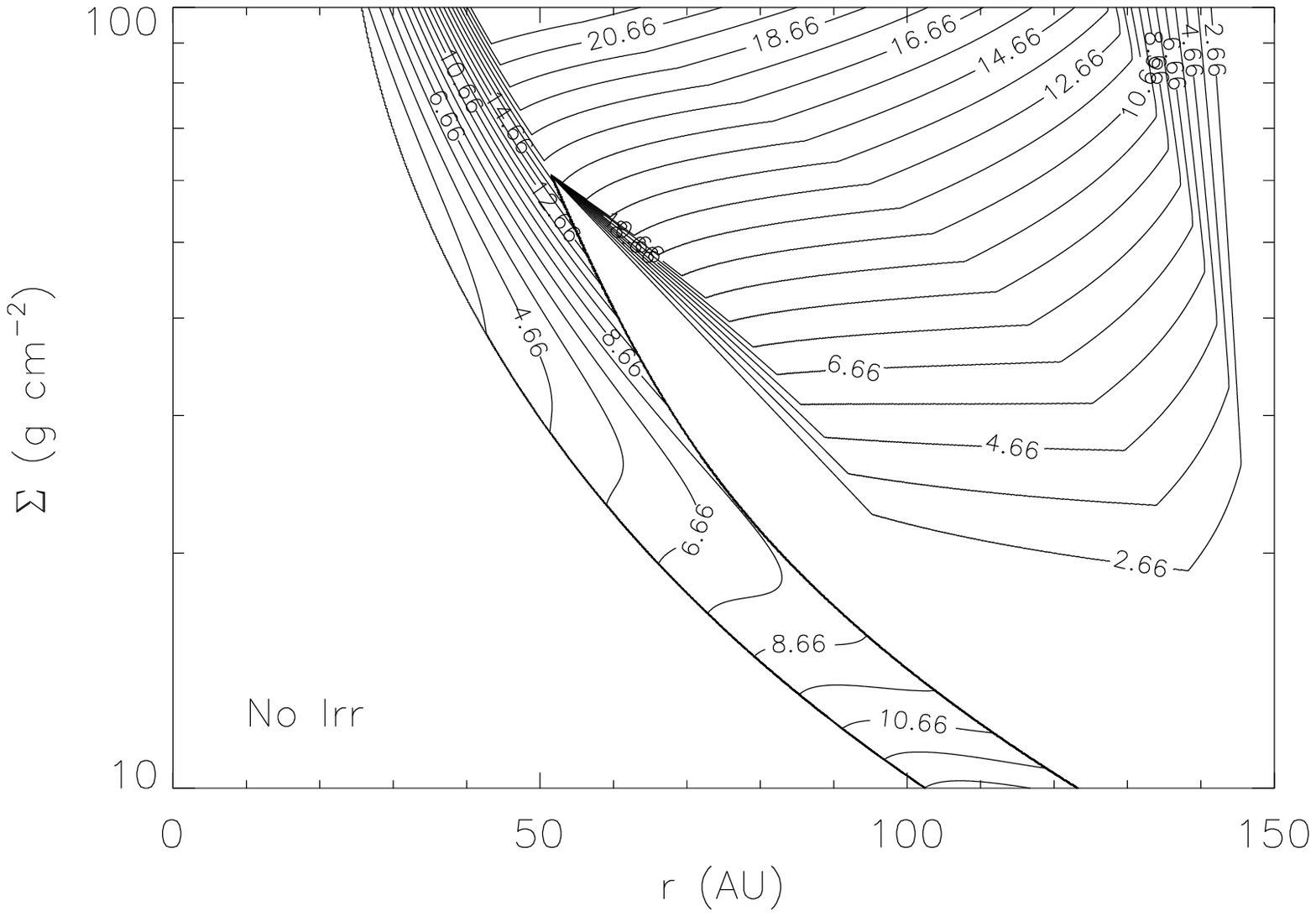} &
\includegraphics[scale = 0.5]{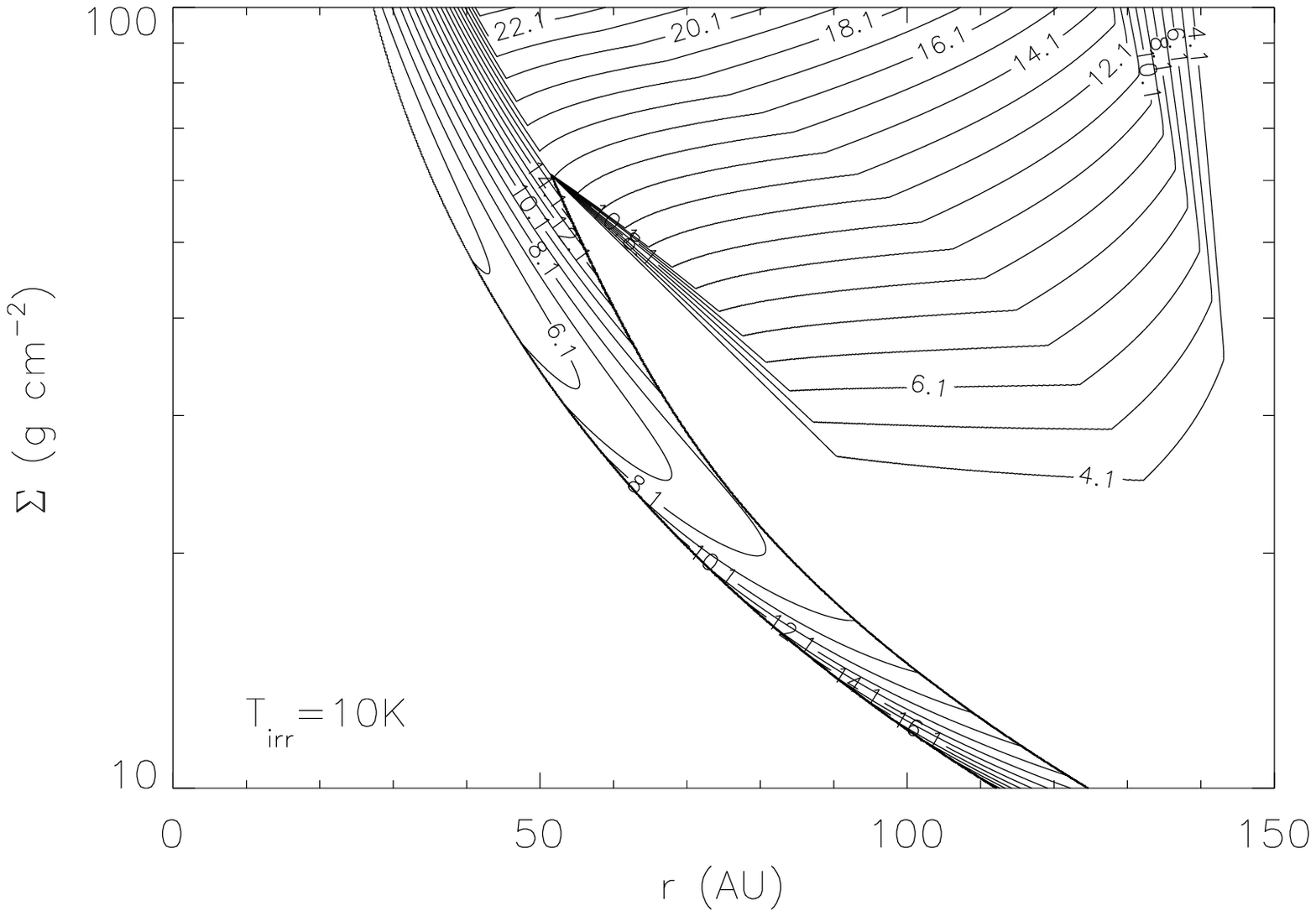} \\
\includegraphics[scale = 0.5]{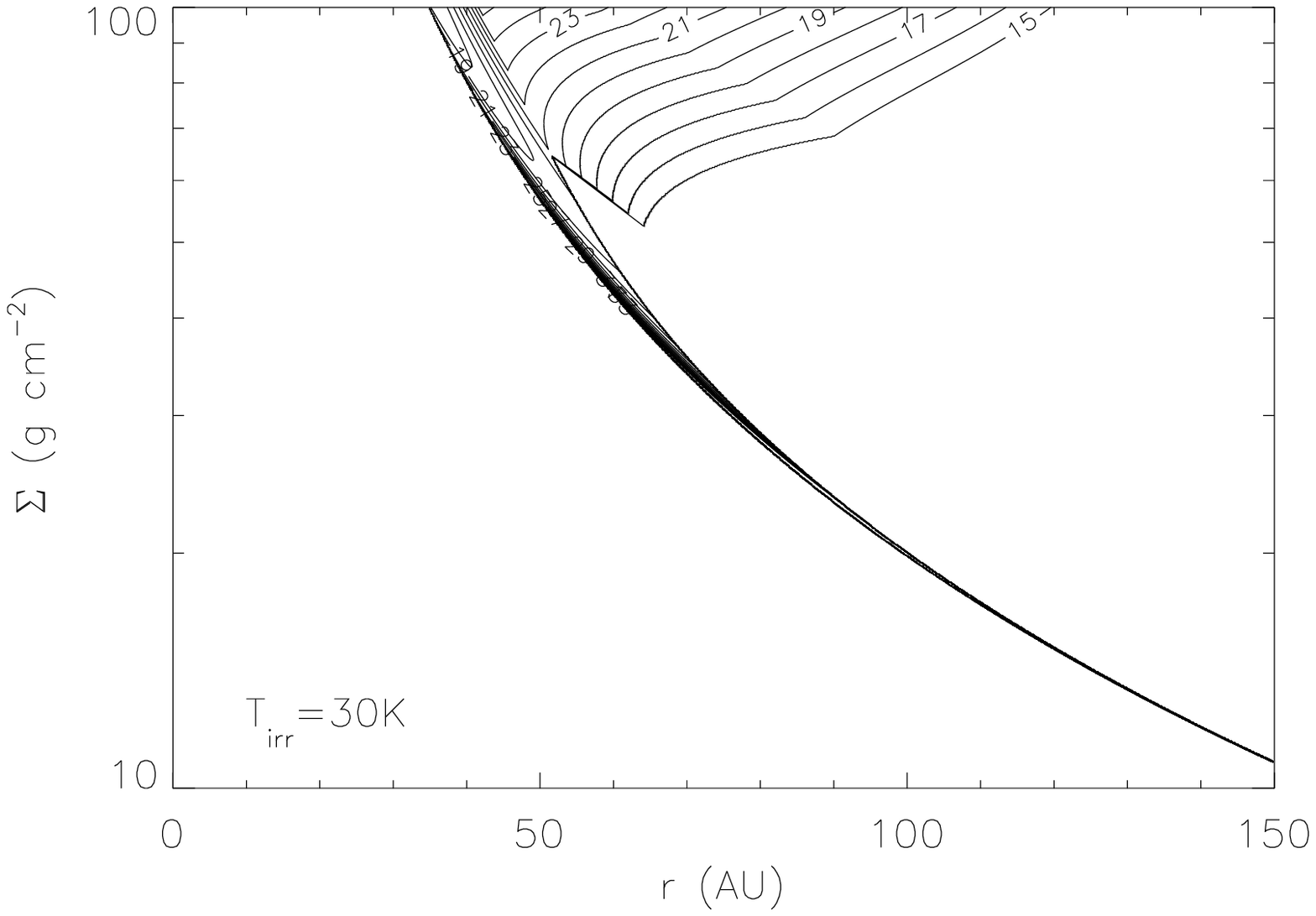} &
\includegraphics[scale = 0.5]{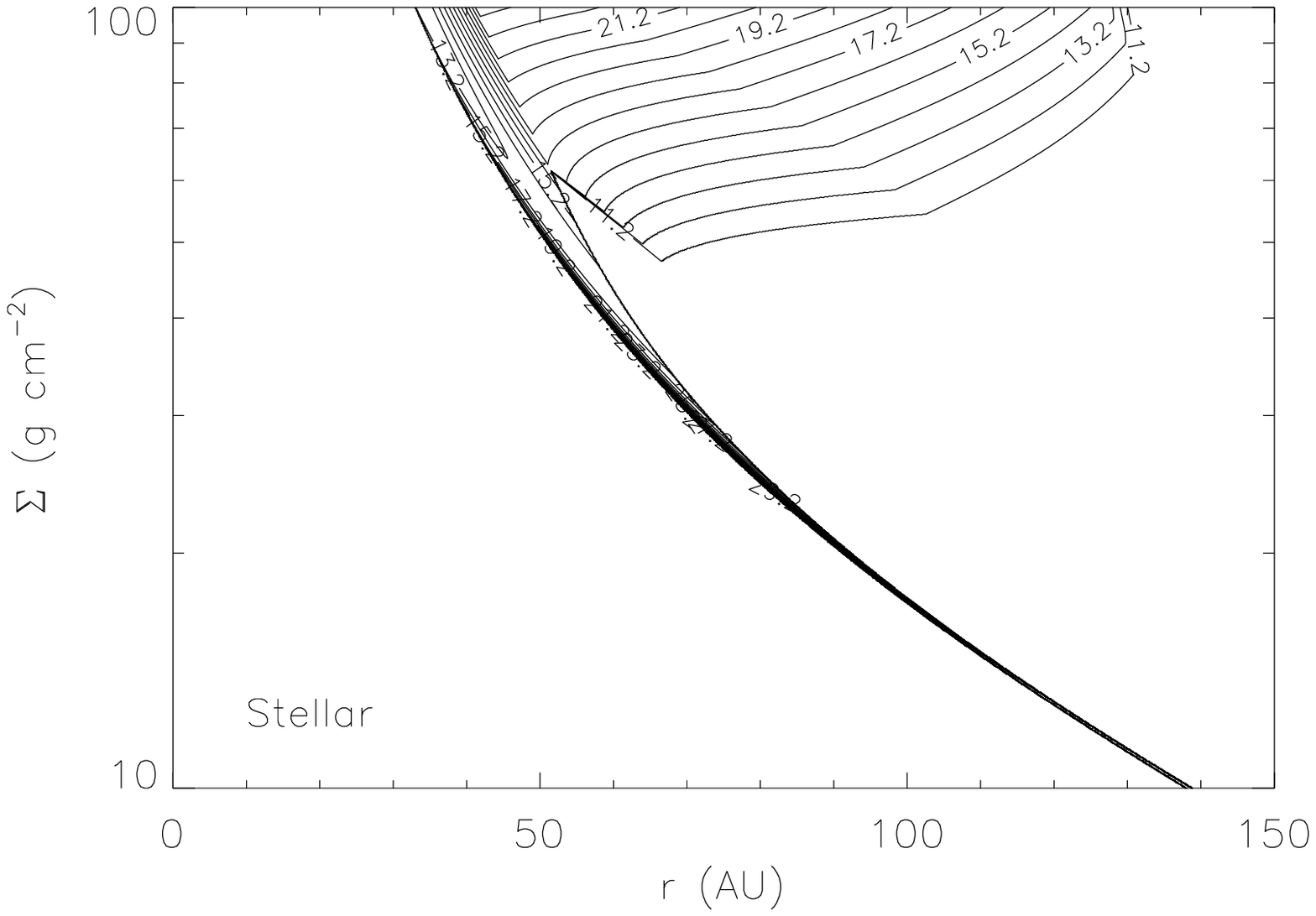} \\
\end{array}$
\caption{2D contour plots of the Jeans mass inside a spiral
  perturbation for (left) a disc irradiated at a background
  temperature of 10 K, and (right) irradiation from the central star,
  as a function of the disc surface density $\Sigma$ and radius in the
  disc $r$, for the four test cases considered in this work: without
  irradiation (top left), with background irradiation at 10K (top
  right), with background irradiation at 30K (bottom left), and with
  stellar irradiation (bottom right).  Regions without contours do not
  possess a fragmenting solution. }\label{fig:sigma_r_frag}
\end{center}
\end{figure*}

\noindent We have established irradiation suppresses fragmentation at
lower accretion rates, and boosts disc masses.  If this is the case,
the surface density at the point of fragmentation should typically be
higher.  In Figure \ref{fig:sigma_r_frag}, we again plot contours
showing the Jeans mass of the fragments, but now study the $\Sigma-r$
parameter space. As expected, lower mass fragments form at lower
surface densities, and the fragment mass tends to increase with
$\Sigma$.  The disc lifetime criterion described in the previous
section excludes part of the parameter space, increasing the gap seen
in the contours from approximately 50 au onwards.

We see that irradiation does not affect the general trend at high
$\Sigma$, but the low $\Sigma$ fragmentation is suppressed.  This is
seen weakly in the background irradiation case, and much more strongly
in the stellar irradiation case.  Low surface density regions will
generally be optically thin, and hence have their temperatures set by
irradiation, remaining too warm to produce fragments.

\subsection{Cumulative Distributions}

\begin{figure*}
\begin{center}$
\begin{array}{cc}
\includegraphics[scale = 0.5]{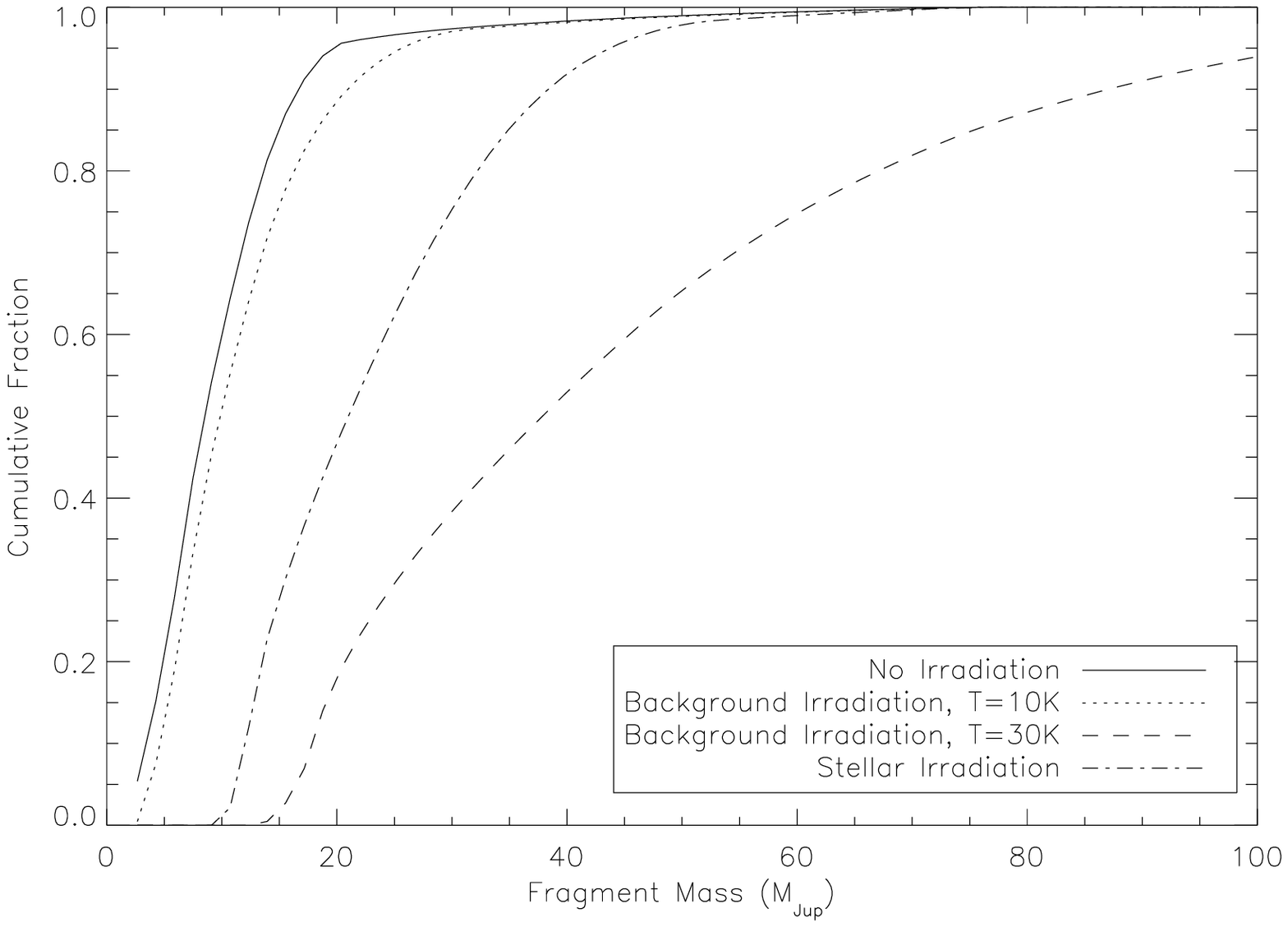} &
\includegraphics[scale=0.5]{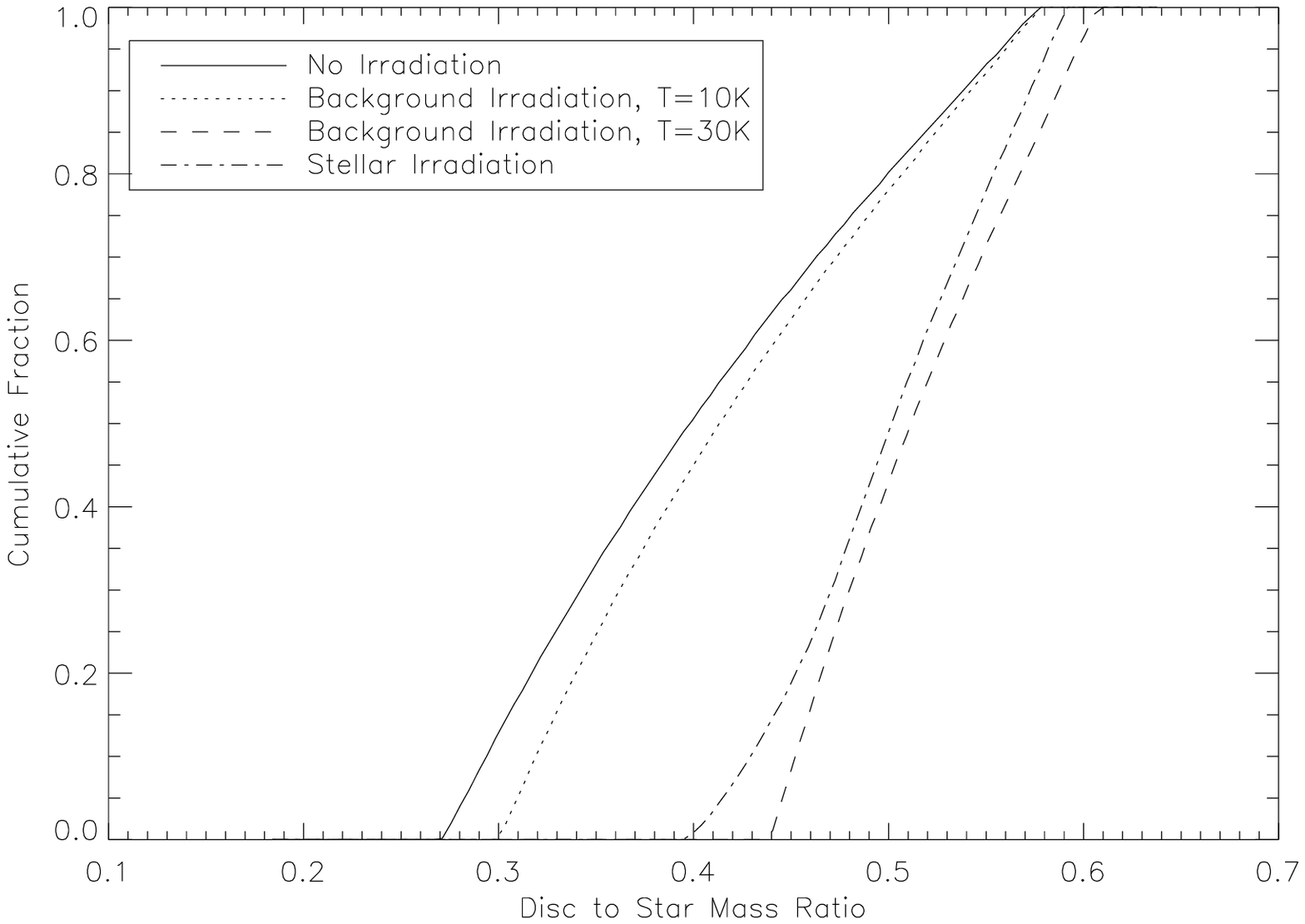} \\
\end{array}$
\caption{Left:The cumulative fraction of fragment mass in the
  parameter space investigated, assuming all values of $\dot{M}$
    and $r_{\rm out}$ are equally likely to occur. Right: The
  cumulative fraction of disc-to-star mass ratio for all models which
  produced fragments (assuming all discs extended to the maximum
  radius of 150 au).}\label{fig:bin}
\end{center}
\end{figure*}

\noindent Although a uniform parameter space in $\dot{M}$ and $r$ is
not what nature presents to us, it is instructive to see the resulting
probability distributions for fragment mass (and disc-to-star mass
ratio) that can be inferred \emph{if we assume that each $\dot{M}-r$
  combination in the parameter space is equally likely}.  These can
then be convolved with the true $(\dot{M},r)$ distributions (once
known) to predict the observed distributions of fragment properties.

The left panel of Figure \ref{fig:bin} shows the cumulative
distribution of fragment masses found in this uniform parameter
study. As low $\Sigma$ regions do not fragment when irradiated, and
disc masses are boosted, consequently the initial fragment masses are
typically higher for the background and stellar irradiation cases
(where the effect becomes significant for the stellar case).  The peak
of the mass distribution is shifted upward by around a factor of 2
(from around 10 $M_{\rm Jup}$ to around 20 $M_{\rm Jup}$) compared to
the case without irradiation.

With the surface density at fragmentation higher in the presence of
irradiation, and the disc masses typically higher for a given
$\dot{M}-r$ locus, one might assume the typical disc-to-star mass
ratio would itself be higher. In the right panel of Figure
\ref{fig:bin}, we plot the cumulative distribution of disc-to-star
mass ratios, assuming that all fragmenting discs extend to $r=150$ au
so as to make the comparisons consistent.

The resulting distributions depend on what fraction of the $\dot{M}-r$
parameter space permits fragmentation. From Figure
\ref{fig:mdot_r_frag}, we can see the non-irradiated case permits
fragmentation in the largest area of parameter space, and at the
lowest accretion rates.  As such, the disc masses required for
fragmentation are the lowest, beginning at around 0.2 $M_{\rm \odot}$
(for the selected disc outer radius of 150 au).  The 10K irradiation
case fragments at slightly higher $\dot{M}$, and boosts disc mass
slightly for the same accretion rate, and as such shows a slightly
higher cumulative fraction.  The $30K$ case fragments at noticeably
higher accretion rates, as does the stellar irradiation case.
However, stellar irradiation suppresses fragmentation at large radii
and low accretion rates, preventing lower disc masses from
fragmenting.  This does not occur for the 30K case, and as such its
distribution is skewed towards even higher masses than the stellar
irradiation case.

\section{Conclusions }\label{sec:conclusions}

\noindent We have continued our investigation of the Jeans mass in
spiral arm perturbations as a means of predicting disc fragmentation
and fragment masses \citep{Forgan2011a}, extending the formalism to
include the effects of irradiation, either from a constant background
or from stellar irradiation.  We construct simple one dimensional disc
models, where the disc maintains thermal equilibrium between the
heating due to self-gravitating spiral structures, irradiation and the
radiative cooling.  From this, we can discover the region of
$\dot{M}-r$ parameter space in which fragmentation is expected to
occur.  We can also determine the expected fragment mass as a function
of disc parameters.  We compare three scenarios: non-irradiated
self-gravitating discs, self-gravitating discs in a uniform
temperature bath, and self-gravitating discs irradiated by the central
object.

Adding irradiation alters the equilibrium disc mass for a given
accretion rate and outer radius, requiring higher mass
self-gravitating discs to exist for these parameters.  We find that in
general, adding irradiation will suppress fragmentation at low
accretion rates compared to the non-irradiated case.  This is a
similar conclusion to that drawn by \citet{Rafikov2009}.  This is not
surprising, as the analysis used here is similar, the only difference
being the precise application of the maximum stress condition for
fragmentation.

Irradiated discs typically must sustain a higher accretion rate in
order to fragment, and the local surface density at fragmentation is
typically higher, as low surface density regions will not be
gravitationally unstable due to the radiative heating.  The fragment
mass is typically boosted, and, as a result, disc fragmentation is
more likely to form brown dwarfs and/or low mass stars than gas giant
planets (unless some action of the disc or central star can prevent
the fragments growing to their isolation mass).

As this work investigates azimuthally averaged disc profiles, we are
not able to address the possibility of irradiated stochastic
fragmentation (cf \citealt{Paardekooper2012}), which suggests regions
which experience a small density perturbation can be irradiated in
such a fashion that the local cooling time is reduced, and
fragmentation could subsequently occur.  This may be further assisted
by the opacity regime in which the perturbation resides
\citep{Cossins2010}.

This stochastic fragmentation condition is equivalent to a local
density perturbation having a reduced Jeans mass for a sufficient
duration to allow self-gravitating collapse to produce a bound object.
The key issue here is: can the local value of $Q$ also remain low? As
we fix $Q$ when carrying out these models, we cannot answer this
question.  Future work should investigate the behaviour of the Jeans
mass in numerical simulations, both with and without irradiation, to
firstly establish how the Jeans mass evolves during a ``realistic''
fragmentation event, and also to test whether stochastic fragmentation
is a sensible outcome of Jeans mass evolution.

To some extent, we therefore agree with \citet{Kratter2011}:
irradiated discs that possess high accretion rates can still be driven
to fragmentation, i.e. more massive discs are required with sufficient
infall to maintain them, as well as high surface densities and
temperatures sufficiently low that gravitational instability can still
arise.  We would also agree with \citet{Stamatellos2008} that the last
of these three conditions is likely to be the greatest obstacle to the
fragmentation of irradiated self-gravitating discs.

\section*{Acknowledgments}

\noindent DF and KR gratefully acknowledge support from STFC grant ST/J001422/1.

\bibliographystyle{mn2e} 
\bibliography{jeans_irr}

\appendix

\label{lastpage}

\end{document}